\documentclass[useAMS,usenatbib]{mn2e}

\usepackage{amssymb}
\usepackage{graphicx} 
\title[Structure formation with scalar field dark matter: the fluid approach]{Structure formation with scalar field dark matter: the fluid approach}
\author[A. Su\'arez and T. Matos]{A. Su\'arez$^{1}$\thanks{E-mail:asuarez@fis.cinvestav.mx ; tmatos@fis.cinvestav.mx} and T. Matos          $^{1}$\\
$^{1}$Departamento de F\'isica, Centro de Investigaci\'on y de Estudios Avanzados del IPN, 07000 M\'exico D.F., M\'exico}
\begin{document}



\maketitle

\label{firstpage}

\begin{abstract}
The properties of nearby galaxies that can be observed in great detail suggest that a better theory rather than cold dark matter (CDM) would 
describe in a better way a mechanism by which matter is more rapidly gathered into large-scale structure such as galaxies and groups of 
galaxies. In this work we develop and simulate a hydrodynamical approach for the early formation of structure in the Universe, this 
approach is based on the fact that dark matter is on the form of some kind of scalar field (SF) with a potential that goes as $\mu^2\Phi^2/2+\lambda\Phi^4/4$, we expect that the fluctuations coming from the SF will give us some information about the matter distribution we observe these days.
\end{abstract}


\section{Introduction}

We begin this work remebering the framework of the standard cosmological model: a homogeneous and isotropic Universe whose evolution is best described by Friedmann's equations that come from general relativity and whose main ingridients can be described by fluids whose characteristics are very similar to those we see in our Universe. Of course, the Universe is not exactly homogeneous and isotropic but this standard model does give us a framework within which we can study the evolution of structure like the observed galaxies or clusters of galaxies from small fluctuations in the density of the early Universe. In this model 4 per cent of the mass in the Universe is in the baryons, 22 per cent is non-baryonic dark matter and the rest in some form of cosmological constant. Another idea that has been around just a bit less than hundred years and in which many of the cosmological models are based in is that of an homogeneous and isotropic Universe, although it has always been clear that this homogeneity and isotropy are only found until certain level. Now we know that the anisotropies are very important and can grow as big as the large scale structure we see today.

 Nowadays the most accepted model in cosmology which explains the evolution of the Universe is known as 
$\Lambda$CDM, because it has achieved some observations with outstanding success, like for example the fact that the cosmic microwave background can be explained in great detail and that it provides a framework within one can understand the large-scale isotropy of the Universe and important characteristics on the origin, nature and evolution of density fluctuations which are believed to give rise to galaxies and other cosmic structure. There remain, however, certain problems at galactic scales, like the cusp profile of central densities in galactic halos, the over 500 substructures predicted by numerical simulations which are not found in observations, etc.. See for example \cite{b16}, \cite{b4} and \cite{b15}.

 In the big bang model, gravity plays an essential role, it collects the dark matter in concentrated regions called 'Dark matter 
haloes'. In the large dark matter haloes, the baryons are believed to be so dense as to radiate enough energy so they will collapse into galaxies and stars. The most massive haloes, hosts for the brightest galaxies, are formed in regions with the highest local mass density. Less massive haloes, hosts for the less bright galaxies, appear in regions with low local densities, i.e, regions were the local density is not well defined, \cite{b14}. These situations appear to be the same as in our extragalactic neighborhood, but still there are problems.

 Observations point out to a better understanding of the theory beginning with the less occupied space called the 'Local void', 
which contains just a few galaxies which are bigger than the expected. This problem would be solved, if the structure grew faster than it does in the standard theory, therefore filling the local void and giving rise to more matter in the surroundings, \cite{b14}.

 Another problem arises for the so called 'Pure disk galaxies', which do not appear in numerical simulations of structure formation in the 
standard theory, because it is believed that their formation which is relatively slow began in the thick stellar bulges. Again this problem would be solved for the early formation of structure.

 The incorporation of a new kind of dark matter, different from the one proposed by the $\Lambda$CDM model into the big bang theory holds 
out the possibility of resolving some of these issues.

 Recent works have introduced a dynamic scalar field with a certain potential $V(\Phi)$ as a candidate to dark matter, although there is not 
yet an agreement for the correct form of the potential of the field. \cite{b20}, and independently \cite{b12}  suggested bosonic dark matter (SFDM) as a model for galactic halos. Another interesting work pointing this way was done by \cite{b13} and independently by \cite{b21} where they used a potential of the form $\cosh$ to explain the core density problem for disc galaxy halos in the $\Lambda$CDM model. \cite{b12} presented a model for the dark matter in spiral galaxies, in which they supposed that dark matter is a scalar field endowed with a scalar potential.

 Several recent work have also suggested that SFDM can be composed of spin-0 bosons which give rise to Bose-Einstein Condensates (BECs), 
which at the same time can make up the galaxies we are observing in our Universe. \cite{b6} proposed that dark matter is composed of ultra/light scalar particles who are initially in the form of a BEC. In their work \cite{b18} used a bosonic dark matter model to explain the structure formation via high-resolution simulations, finally \cite{b17} reviewed the key properties that may arise from the bosonic nature of SFDM models.

 The main objective of this work in difference with others is to introduce SFDM and assume that dark matter its itself a scalar field that 
involves an auto-interacting potential of the form $V(\Phi)=\mu^2\Phi^2/2+\lambda\Phi^4/4$, where $\mu_{\Phi}\sim 10^{-22}$ eV is the mass of the scalar field, \cite{b20}, \cite{b10} and \cite{b6}. With the mass $\mu_{\Phi} \sim 10^{-22}$ eV and only one free parameter when $\lambda$ is taken equals to zero, the SFDM model fits the following important features:
 \begin{enumerate}
  \item The cosmological evolution of the density parameters of all the components of the Universe, \cite{b9}.
  \item The rotation curves of galaxies, \cite{b3}, and the central density profile of LSB galaxies, \cite{b2}, 
  \item With this mass, the critical mass of collapse for a real scalar field is just $10^{12}\,M_{\odot}$, i.e., the one observed in 
        galactic haloes, \cite{b1}.
  \item The central density profile of the dark matter is flat, \cite{b2}. 
  \item The scalar field has a natural cut off, thus the substructures in clusters of galaxies is avoided naturally. With a scalar field 
   mass of $\mu_\Phi\sim10^{-22}$ eV the amount of substructures is compatible with the observed one, \cite{b10}.
 \end{enumerate}
 In this paper we show that the SFDM predicts galaxy formation earlier than the cold dark matter model, because they form BEC at a critical 
temperature $T_c >>$ TeV. So, if SFDM is right, this would imply that we have to see big galaxies at high redshifts. In order to do this, we study the density fluctuations of the scalar field from a hydrodynamical point of view, this will give us some information about the energy density of dark matter halos necessary to obtain the observational results of large-scale structure. Here we will give some tools that might be necessary for the study of the early formation of structure.

 In section \ref{fondo} we analyse the analytical evolution of the SF, then in section \ref{fluct} we treat the SF as a hydrodynamical fluid 
in order to study its evolution for the density contrast, in section \ref{results} we compare our results with those obtained by the CDM model for the density contrast in the radiation dominated era just before recombination and finally we give our conclusions.

\section[]{The Background}\label{fondo}

In this section we perform a transformation in order to solve the Friedmann equations analytically with the approximation $H<<\mu_{\Phi}$. The scalar field (SF) we deal with depends only on time, $\Phi=\Phi_0(t)$, and of course the background is only time dependent as well.

 We use the Friedmann-Lema\^itre-Robertson-Walker (FLRW) metric with scale factor $a(t)$. The background Universe is composed only by
SFDM ($\Phi_{0}$) endowed with a scalar potential. We begin by recalling the basic background equations. From the energy-momentum tensor $\mathbf{T}$ for a scalar field, the scalar energy density $T_0^0$ and the scalar pressure $T_j^i$ are given by
 \begin{equation}
  T_0^0=-\rho_{\Phi_0}=-\left(\frac{1}{2}\dot{\Phi}_0^2+ V\right),
 \label{rhophi0}
 \end{equation}
 \begin{equation}
  T_j^i=p_{\Phi_0}=\left(\frac{1}{2}\dot{\Phi}_0^2-V\right)\delta_j^i,
 \label{pphi0}
 \end{equation}
where the dots stand for the derivative with respect to the cosmological time and $\delta^i_j$ is Kronecker's delta. Thus, the Equation of State (EoS) for the scalar field is $p_{\Phi_{0}}=\omega_{\Phi_{0}}\,\rho_{\Phi_0}$ with
 \begin{equation}
  \omega_{\Phi_{0}}=\frac{\frac{1}{2}\dot{\Phi}_{0}^{2}\,-\,V}{\frac{1}{2}\dot{\Phi}_{0}^{2}\,+\,V}.
 \label{ec:w}
 \end{equation}

 Notice that background scalar quantities have the subscript $0$. Now the following dimensionless variables are defined
 \begin{eqnarray}
  x\equiv\frac{\kappa}{\sqrt{6}}\frac{\dot{\Phi}_{0}}{H},\quad 
  u\equiv\frac{\kappa}{\sqrt{3}}\frac{\sqrt{V}}{H},\nonumber
 \label{eq:varb}
 \end{eqnarray}
being $\kappa^{2} \equiv 8\pi G$ and $H \equiv \dot{a}/a$ the Hubble parameter. Here we take the scalar potential as $V=m^{2}\Phi^2/2\hbar^2+\lambda\Phi^4/4$, where, $\mu=mc/\hbar$ and $m$ is the mass given in kilograms, and from now on we will use units where $c=1$, then for the ultra-light boson particle we have that $\mu_{\Phi}\sim 10^{-22}$ eV.

 With these variables, the density parameter $\Omega_\Phi$ for the background $0$ can be written as
 \begin{eqnarray}
  \Omega_{\Phi_{0}}=x^2+u^2.
 \label{eq:dens}
 \end{eqnarray}
In addition, we may write the EoS of the scalar field as
 \begin{equation}
  \omega_{\Phi_{0}}=\frac{x^{2}-u^{2}}{\Omega_{\Phi_{0}}}.
 \label{eq:dlw}
 \end{equation}
Since $\omega_{\Phi_{0}}$ is a function of time, if its time average tends to zero, this would imply that $\Phi^2$-dark matter can be able to mimic the EoS for CDM, see \cite{b11} and \cite{b9}.

Now we express the SF, $\Phi_0$, in terms of the new variables $S$ and $\hat\rho_0$, where $S$ is constant in the background and $\hat{\rho}_0$ will be the energy density of the fluid also in the background. So, our background field is proposed as
 \begin{equation}
  \Phi_0=(\psi_0\rmn e^{-\rmn imt/\hbar}+\psi^*_0\rmn e^{\rmn imt/\hbar})
 \end{equation}
where,
 \begin{equation}
  \psi_0(t)=\sqrt{\hat{\rho}_0(t)}\rmn e^{\rmn iS/\hbar}
 \end{equation}
and with this our SF in the background can be finally expressed as,
 \begin{equation}
  \Phi_0=2\sqrt{\hat{\rho}_0}\cos(S-mt/\hbar),
 \label{tri}
 \end{equation}
with this we obtain
 \begin{eqnarray}
  \dot{\Phi}_0^2&=&\hat{\rho}_0\left[\frac{\dot{\hat{\rho}}_0}{\hat{\rho}_0}\cos(S-mt/\hbar)\right.\nonumber\\
  &-&\left.2(\dot{S}-m/\hbar)\,\sin(S-mt/\hbar)\right]^2
 \label{backSF}
 \end{eqnarray}

 To simplify, observe that the uncertanty relation implies that $m\Delta t\sim\hbar$, and for the background in the non-relativistic case 
the relation $\dot S/m\sim0$ is satisfied. Notice also that for the background we have that the density goes as $(\ln\hat{\rho}_0)\dot{}=-3H,$ but we also have that $H\sim 10^{-33}$ eV $<<\mu_{\Phi}\sim 10^{-22}$ eV, so with these considerations at hand for the background, in (\ref{backSF}) we have
 \begin{equation}
  \dot{\Phi}_0^2=4\frac{m^2}{\hbar^2}\hat{\rho}_0\sin^2(S-mt/\hbar)
 \end{equation}
 
Finally, substituting this last equation and equation (\ref{tri}) into (\ref{rhophi0}) when taking $\lambda=0$, we obtain
  \begin{equation}
   \rho_{\Phi_0}=2\frac{m^2}{\hbar^2}\hat{\rho}_0[\sin^2(S-mt/\hbar)+\cos^2(S-mt/\hbar)]=2\frac{m^2}{\hbar^2}\hat{\rho}_0.
  \label{trigo}
  \end{equation}

 Comparing this result with (\ref{eq:dens}) we have that the identity $\Omega_{\Phi_0}=2m^2\hat{\rho}_0/\hbar^2$ holds for the background, 
so comparing with (\ref{trigo}),
 \begin{equation}
  x=\sqrt{2\hat{\rho}_0}\frac{m}{\hbar}\sin(S-mt/\hbar)
 \label{eq:cinet}
 \end{equation}
 \begin{equation}
  u=\sqrt{2\hat{\rho}_0}\frac{m}{\hbar}\cos(S-mt/\hbar).
 \label{eq:potencial}
 \end{equation}

\begin{figure}
 \scalebox{0.6}{\includegraphics{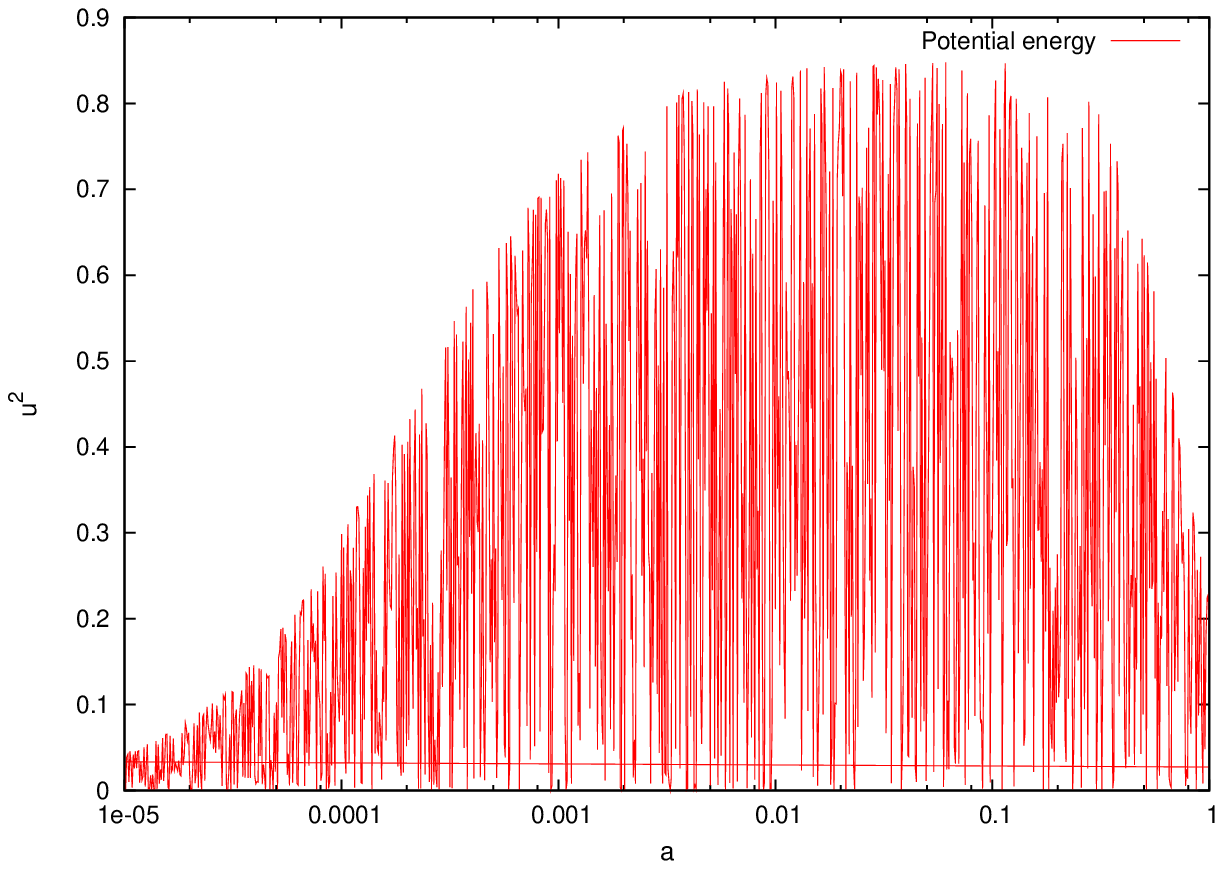}}
 \scalebox{0.6}{\includegraphics{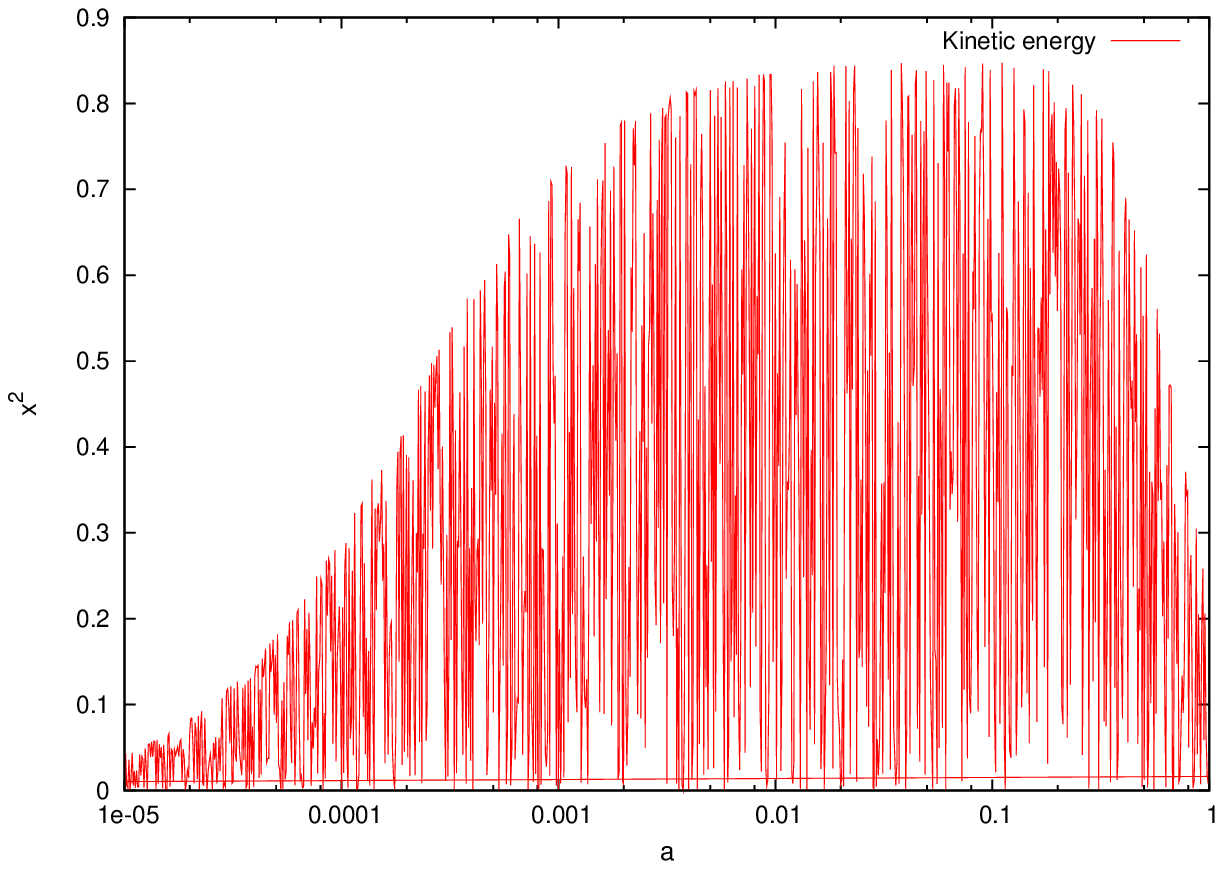}}
 \caption{Analytical evolution of the potential (top panel) and kinetic (bottom panel) energies of the scalar field dark matter.}
 \label{fig1}
 \end{figure}

 We plot the evolution of the energies (\ref{eq:cinet}) and (\ref{eq:potencial}) in Fig. \ref{fig1}, where for the evolution we 
used the e-folding number $N$ defined as $N=ln(a)$ and the fact that $a\sim t^n\rightarrow t\sim\rmn{e}^{N/n}.$ In terms of the two analytic results (\ref{eq:cinet}) and (\ref{eq:potencial}) Fig. \ref{fig1} shows the kinetic and the potential energies of the scalar field.
Observe the excelent accordance with the numerical results in \cite{b9} for the kinetic and potential energies of the background respectively.

\section{Scalar Field Fluctuations}\label{fluct}

 If dark matter is some kind of elemental particle with mass $\mu$, then it would be about $10^{68}\mu$ GeV$^{-1}$ particles to follow in a 
single galaxy.

 Here we describe a model for the non-interacting matter such that: i) It can describe it more as a field than as particles and ii) We find 
a function that only depends on the three spatial coordinates and time.

 Now a days it is known that our Universe is not exactly isotropic and spatially homogeneous like the FLRW metric describes. There exist 
small deviations from this model, and if we believe these deviations are small enough, they can be treated by the linear perturbation theory.

 Then, if dark matter is composed of scalar particles with masses $\mu<<1$ eV, the occupation numbers in galactic haloes are so big that the 
dark matter behaves as a classical field that obeys the Klein-Gordon equation $(\square^2+m^2/\hbar^2)\Phi=0$, where $\Box$ is the D'Alambertian and we have set $c=1$.

 By definition, a perturbation done in any quantity, is the difference between its value in some event in real space-time, and its 
corresponding value in the background. So, for example for the SF we have
 \begin{equation}
  \Phi=\Phi_0(t)+\delta\Phi(\mbox{\boldmath$x$},t),
 \label{pert}
 \end{equation}
where the background is only time dependent, while the perturbations also depend on the space coordinates. Similar cases apply for the metric;
 \begin{eqnarray}
  g_{00}&=&-a^2(1+2\phi),\nonumber\\
  g_{0i}&=&a^2B,_i,\nonumber\\
  g_{ij}&=&a^2[(1-2\psi)\delta_{ij}+2E,_{ij}].
 \end{eqnarray}
Here the scale factor $a$ depends on the conformal time, $\psi$ is a perturbation associated to the curvature and E is asociated to the expansion. We will work under the Newtonian gauge, which is defined when $B=E=0$. An advantage of using this gauge is that here the metric tensor $g_{\mu \nu}$ is diagonal, and so the calculations become much easier. We will only work with scalar perturbations, vector and tensor perturbations are eliminated from the beginning, so that only scalar perturbations are taken into account. Another advantage in using this gauge is that $\phi$ will play the role of the gravitational potential, this will help us to have a simpler physical interpretation, i.e., both potentials $\phi$ and $\psi$ are then related. This metric has already been used in other works, \cite{b23}, \cite{b19} and \cite{b8}.

 For the perturbed Klein-Gordon where we have used equation (\ref{pert}) and we have set $\dot{\phi}=0$, we have:
 \begin{equation}
  \delta\ddot{\Phi}+3H\delta\dot{\Phi}-\frac{1}{a^2}\hat\nabla^2\delta\Phi+V,_{\Phi\Phi}\delta\Phi+2V,_{\Phi}\phi=0
 \label{K-G}
 \end{equation}

 The SF $\Phi$ has very hard oscillations from the beginning, this oscillations are transmitted to the fluctuations which apparently seems 
to grow very fast and are too big. Nevertheless, this behavior is not physical, because we only see the oscillations of the fields, but we can not see clearly the evolution of its density, \cite{b11}. In order to drop out these oscillations, in what follows we perform two transformations. The first one changes the perturbed Klein-Gordon equation into a kind of 'Schr\"odinger' equation and the second transforms this last equation into a hydrodynamical system, where we can interpret the physical quantities easier and the observable quantities become much clear. Now we express the perturbed SF $\delta\Phi$ in terms of the field $\Psi$,
 \begin{equation}
  \delta\Phi=\Psi \rmn{e}^{-\rmn{i}mt/\hbar}+\Psi^*\rmn{e}^{\rmn{i}mt/\hbar},
 \end{equation}
term which oscillates with a frequency proportional to $m$ and $\Psi=\Psi(\mbox{\boldmath$x$},t)$ which would be proportional to a wave function of an ensamble of particles in the condensate. With this equation and the expresion for the potential of the scalar field, (\ref{K-G}) transforms into
   \begin{equation}
    -\rmn{i}\hbar(\dot{\Psi}+\frac{3}{2}H\Psi)+\frac{\hbar^2}{2m}(\square\Psi+9\lambda|\Psi|^2\Psi)+m\phi\Psi=0,
   \label{schrodinger}
   \end{equation}
where we have defined
 \begin{equation}
  \square=\frac{\rmn{d}^2}{\rmn{d}t^2}+3H\frac{\rmn{d}}{\rmn{d}t}-\frac{1}{a^2}\hat\nabla^2.
 \end{equation}
Notice that this last equation could represent a kind of 'Gross-Pitaevskii' equation in an expanding Universe. The only modification of equation (\ref{schrodinger}) in comparison to the Schr\"odinger or the Gross-Pitaevskii equation is the scale factor $a^{-1}$ associated to the co-moving spatial gradient and that the Laplacian $\hat\nabla^2=\partial^2_{\mbox{\boldmath{x}}}$ transforms into the D' Alambertian $\Box$.

 To explore the hydrodynamical nature of bosonic dark matter, we will use a modified fluid approach. Then, to make the connection between 
the theory of the field and the condensates waves function, the field is proposed as,
 \begin{equation}
  \Psi=\sqrt{\hat{\rho}}\,\rmn{e}^{\rmn{i}S},
 \label{psi}
 \end{equation}
where $\Psi$ will be the condensates wave function with $\hat{\rho}=\hat{\rho}(\mbox{\boldmath$x$},t)$ and $S=S(\mbox{\boldmath$x$},t)$, \cite{b5}. Here we have separated $\Psi$ into a real phase $S$ and a real amplitude $\sqrt{\hat{\rho}}$ and the condition $\mid\Psi\mid^2=\Psi\Psi^*=\hat{\rho}$ is satisfied. From (\ref{psi}) we have
 \begin{eqnarray}
  \dot{\hat{\rho}}&+&3H\hat{\rho}-\frac{\hbar}{m}\hat{\rho}\Box S+\frac{\hbar}{a^2m}\hat\nabla S\hat\nabla\hat{\rho}
   -\frac{\hbar}{m}\dot{\hat{\rho}}\dot{S}=0\nonumber\\
  \hbar\dot S/m&+&\omega\hat{\rho}+\phi+\frac{\hbar^2}{2m^2}(\frac{\hat\Box\sqrt{\hat{\rho}}}{\sqrt{\hat{\rho}}})
  +\frac{\hbar^2}{2a^2}[\hat\nabla(S/m)]^2\nonumber\\ 
  &-&\frac{\hbar^2}{2}(\dot S/m)^2=0\label{hidro1}
 \label{hidro}
 \end{eqnarray}
Now, taking the gradient of (\ref{hidro1}) then dividing by $a$ and using the definition
 \begin{equation}
  \mbox{\boldmath$v$}\equiv\frac{\hbar}{ma}\hat\nabla S
 \label{vel}
 \end{equation}
we have,
 \begin{eqnarray}
  \dot{\hat{\rho}}&+&3H\hat{\rho}-\frac{\hbar}{m}\hat{\rho}\Box S+\frac{1}{a}\mbox{\boldmath$v$}\nabla\hat{\rho}
   -\frac{\hbar}{m}\dot{\hat{\rho}}\dot{S}=0\nonumber\\
  \dot{\mbox{\boldmath$v$}}&+&H\mbox{\boldmath$v$}+\frac{1}{2a\hat{\rho}}\nabla p+\frac{1}{a}\nabla\phi
  +\frac{\hbar^2}{2m^2a}\nabla(\frac{\Box\sqrt{\hat{\rho}}}{\sqrt{\hat{\rho}}})\nonumber\\
  &+&\frac{1}{a}(\mbox{\boldmath$v\cdot$}\nabla)\mbox{\boldmath$v$}-\hbar(\dot{\mbox{\boldmath$v$}}+H\mbox{\boldmath$v$})
  (\dot S/m)=0\label{eq:1}.
 \end{eqnarray}
where in (\ref{hidro1}) $\omega=9\hbar^2\lambda/2m^2$ and in (\ref{eq:1}) we have defined $p=\omega\hat{\rho}^2$.

  It is worth noting that, to this moment this last set of equations do not involve any approximations with respect to equation 
(\ref{schrodinger}) and can be used in linear and non-linear regimes.

 Now, neglecting squared terms, second order time derivatives and products of time derivatives in this last set of equations we get,
 \begin{eqnarray}
  \frac{\partial\hat{\rho}}{\partial t}&+&\nabla\mbox{\boldmath$\cdot$} (\hat{\rho}\mbox{\boldmath$v$})+3H\hat{\rho}=0\label{navier}\\
  \frac{\partial\mbox{\boldmath$v$}}{\partial t}&+&H\mbox{\boldmath$v$}+(\mbox{\boldmath$v\cdot$}\nabla)\mbox{\boldmath$v$}- 
  \frac{\hbar^2}{2m^2}\nabla(\frac{1}{2\hat{\rho}}\nabla^2\hat{\rho})+\omega\nabla\hat{\rho}\nonumber\\
  &+&\nabla\phi=0\label{navier1}\\
  \nabla^2\phi&=&4\pi G\hat{\rho}\label{poisson}
 \end{eqnarray}
where the equation for the gravitational field is given by Poisson's equation (\ref{poisson}). In these equations we have introduced $\mbox{\boldmath$r$}=a(t)\mbox{\boldmath$x$}$, such that $1/a\hat\nabla=\nabla=\partial_{\mbox{\boldmath{r}}}$.

 Equation (\ref{vel}) shows the proportionality between the gradient of the phase and the velocity of the fluid. Note that {\boldmath$v$} 
can represent the velocity field for the fluid and $\hat{\rho}$ will be the particles density number within the fluid. Also there exists an extra term of third order for the partial derivatives in the waves amplitude which goes as the gradient of $\frac{\hbar^2}{2m^2}\frac{\Box\sqrt{\hat{\rho}}}{\sqrt{\hat{\rho}}}$, this term would result in a sort of 'quantum pressure' that would act against gravity. We remain that $\phi$ represents the gravitational field. These two sets of equations (\ref{navier}) and (\ref{navier1}) would be analogous to Euler's equations of classical 'fluids', with the main difference that there exists a 'quantum part', which we will call $Q$ and will be given by $Q=\frac{\hbar^2}{2m^2}\frac{\square\sqrt{\hat{\rho}}}{\sqrt{\hat{\rho}}}$ which can describe a force or a sort of negative quantum pressure.

 For equation (\ref{navier}) we have that $\hat{\rho}$ will represent the mass density or the particles density number of the fluid, where 
all the particles would have the same mass. Finally, these equations describe the dynamics of a great number of non-interacting identical particles that manifest themselves in the form of a fluid, also equation (\ref{schrodinger}) can describe a great number of non-interacting but self-interacting identical particles in the way of a Bose gas, when the probability density is interpreted as the density number.

 Now, these hydrodynamical equations are a set of complicated non-linear differential equations. To solve them we will restrict ourselves to 
a vecinity of total equilibrium.

 For this let $\hat{\rho}_0$ be the mass density of the fluid in equilibrium, the average velocity 
$\mbox{\boldmath$v$}_0$ will be taken as zero in equilibrium, so we will only have $\mbox{\boldmath$v$}(\mbox{\boldmath$x$},t)$ out of equilibrium. Then, the matter in the Universe will be considered as a hydrodynamical fluid inside an Universe in expansion. This system will then evolve in this Universe and later on they will collapse because of their gravitational attraction.

   Then from (\ref{navier}) for the mass density of the fluid in equilibrium we have,
   \begin{equation}
    \frac{\partial\hat{\rho}_0}{\partial t}+3H\hat{\rho}_0=0,
   \end{equation}
with solutions of the form
   \begin{equation}
    \hat{\rho}_0=\frac{\rho_{0i}}{a^3},
   \label{back}
   \end{equation}
where as we know, in general if we have an equation of state of the form $\hat p=\omega\hat\rho$ and consider CDM or dust as dark matter such that $\hat p=0$ it holds that $\hat\rho\propto a^{-3}.$ Then, when the scale factor was small, the densities were necessarily bigger. Now, the particles density number are inversely proportional to the volume, and must be proportional to $a^{-3}$, therefore the matter energy density will also be proportional to $a^{-3}$, result that is consistent with our expression (\ref{back}).

Now for the system out of equilibrium we have
 \begin{eqnarray}
  \frac{\partial\delta\hat{\rho}}{\partial t}&+&3H\delta\hat{\rho} 
  +\hat{\rho}_0\nabla\mbox{\boldmath$\cdot$}\delta\mbox{\boldmath$v$}=0\nonumber\\
  \frac{\partial\delta\mbox{\boldmath$v$}}{\partial t}&+&H\delta\mbox{\boldmath$v$} 
  -\frac{\hbar^2}{2m^2}\nabla(\frac{1}{2}\nabla^2\frac{\delta\hat{\rho}}{\hat{\rho}_0})+\omega\nabla\delta\hat{\rho}
  +\nabla\delta\phi=0\nonumber\\
  \nabla^2\delta\phi&=&4\pi G\delta\hat{\rho}
  \label{outeq}
  \end{eqnarray}
equations that are valid in a Universe in expansion. In order to solve system (\ref{outeq}) we look for solutions in the form of plane waves, for this the convenient ansatz goes as 
 \begin{eqnarray}
  \delta{\hat\rho}&=&\hat\rho_1(t)\exp(\rmn{i}\mbox{\boldmath$k\cdot x$}/a),\nonumber\\ 
  \delta\mbox{\boldmath$v$}&=&\mbox{\boldmath$v$}_1(t)\exp(\rmn{i}\mbox{\boldmath$k\cdot x$}/a),\nonumber\\
  \delta\phi&=& \phi_1(t)\exp(\rmn{i}\mbox{\boldmath$k\cdot x$}/a).\nonumber
 \end{eqnarray}
where {\boldmath$x$} is the position vector and {\boldmath$k$} is a real wavevector which corresponds to a wavelength $\lambda$. If we substitute these ansatz in the set of equations (\ref{outeq}), we then have
   \begin{eqnarray}
   \frac{\rmn{d}\hat\rho_1}{\rmn{d} t}&+&3H\hat\rho_1+\rmn{i}\frac{{\hat\rho}_0}{a}\mbox{\boldmath$k\cdot v$}_1=0,\label{eq1}\\
   \frac{\rmn{d}\mbox{\boldmath$v$}_1}{\rmn{d} t}&+&H\mbox{\boldmath$v$}_1+\rmn{i}\frac{\hat\rho_1}{a}\left(\frac{v_\rmn{q}^2}{\hat{\rho}_0}-
     4\pi G\frac{a^2}{k^2}+\omega\right)\mbox{\boldmath$k$}=0,\\
    \phi_1&+&4\pi G\frac{a^2}{k^2}\hat\rho_1=0.
   \label{modos}
   \end{eqnarray}
where we have defined the velocity
\begin{equation}
v^2_\rmn{q}=\frac{\hbar^2k^2}{4a^2m^2}
\label{eq:vq}
\end{equation}
   To solve the system is convenient to rotate the coordinate system so that the propagation of the waves will be along the direction of 
one of the axes. For this we know that the velocity vector can be divided into longitudinal (parallel to {\boldmath$k$}) and transverse (perpendicular to {\boldmath$k$}) parts, so we have $\mbox{\boldmath$v$}_1=\lambda\mbox{\boldmath$k$}+\mbox{\boldmath$v$}_2$, where $\mbox{\boldmath$v$}_2$ is the vector perpendicular to the wave propagation vector $\mbox{\boldmath$k\cdot v$}_2=0$. In terms of $\mbox{\boldmath$v$}_2$ for equations (\ref{eq1})-(\ref{modos}) we have
   \begin{eqnarray}
   \frac{\rmn{d}\hat\rho_1}{\rmn{d} t}&+&3H\hat\rho_1+\rmn{i}\frac{{\hat\rho}_0}{a}k^2\lambda=0\label{eq2}\\
    \frac{\rmn{d}\lambda}{\rmn{d} t}&+&H\lambda+\frac{\rmn{i}}{a}(\frac{v_\rmn{q}^2}{\hat{\rho}_0}-4\pi G\frac{a^2}{k^2}+\omega)\hat\rho_1=0,
    \label{lambda}
   \end{eqnarray}
in addition to an equation for $\mbox{\boldmath$v$}_2$, $\rmn{d}\mbox{\boldmath$v$}_2/\rmn{d} t+H\mbox{\boldmath$v$}_2=0$, with solutions $\mbox{\boldmath$v$}_2=C/a$ with $C$ a constant of integration, i.e., perpendicular modes to the wave vector are eliminated with the expansion of the Universe. Now, if we use the result (\ref{back}), then equation (\ref{eq2}) can be written as
   \begin{equation}
    \frac{\rmn{d}}{\rmn{d}t}\left(\frac{\hat\rho_1}{\hat{\rho}_0}\right)=-\frac{\rmn{i}k^2\lambda}{a}   \label{lambda1}
   \end{equation}

 System (\ref{eq2})-(\ref{lambda}) can be treated as in the case of a Universe with no expansion, so combining the two equations and with 
the aid of (\ref{lambda1}), we get
   \begin{equation}
    \frac{\rmn{d}^2\delta}{\rmn{d}t^2}+2H\frac{\rmn{d}\delta}{\rmn{d}t}+\left[(v_\rmn{q}^2+\omega\hat{\rho}_0)\frac{k^2}{a^2}
    -4\pi G{\rho}_0\right]\delta=0,
   \label{delta}
   \end{equation}
where $\delta=\hat\rho_1/\hat{\rho}_0=\rho_1/{\rho}_0$ is defined as the density contrast. This will be a fundamental equation in the understanding of the evolution of the primordial fluctuations.

\section{Results}\label{results}

First we will give a brief summary of the results for the $\Lambda$CDM model, this will enable us to make a direct comparison with our results.

 For CDM the equation for the evolution of the density contrast is given by,
 \begin{equation}
  \frac{\rmn{d}^2\delta}{\rmn{d}t^2}+2\,H\frac{\rmn{d}\delta}{\rmn{d}t}+\left(c_\rmn{s}^2\frac{k^2}{a^2}-4\pi G\hat{\rho}_0\right)\delta=0,
 \label{deltaCDM}
 \end{equation}
where $c_\rmn{s}$ is defined as the sound velocity (which in our case it is not). Now lets analyze equation (\ref{deltaCDM}) at the beginning of the the matter dominated era a time just after the epoch of equality,and just before recombination when the radiation has cooled down and the photons do not interact with the electrons anymore, for a relativistic treatment see \cite{b50}. In this era, $a\geqslant a_{eq}$, practically all the interesting fluctuation modes are well within the horizon, and the evolution of the perturbations can be well described within the newtonian analysis. At this time, matter behaves like dust with zero pressure. So we have $a\sim t^{2/3}$, $c_\rmn{s}^2k^2/a^2\approx 0$ and $\hat{\rho}_0\sim t^{-2}$ therefore $H=(2/3)1/t$. For equation (\ref{deltaCDM}) we have
 \begin{equation}
  \frac{\rmn{d}^2\delta}{\rmn{d}t^2}+\frac{4}{3}\frac{1}{t}\frac{\rmn{d}\delta}{\rmn{d}t}-\frac{2}{3}\frac{1}{t^2}\delta=0.
 \end{equation}

   The solutions to this equation are of the form
   \begin{equation}
    \delta(t)\to t^{2/3}C_1+\frac{C_2}{t},
   \label{deltat}
   \end{equation}
where $C_1$ and $C_2$ are integration constants, from this solution we can see that we have modes that will disappear as time goes by, and modes that grow proportionally to the expansion of the Universe. This is an important result, because then the density contrast will grow proportionally to the expansion of the Universe when this is dominated by matter. Then, these fluctuations can maybe grow and give life to the galaxies, clusters of galaxies and all the large-scale structure we see now a days.
\begin{figure}
 \scalebox{0.6}{\includegraphics{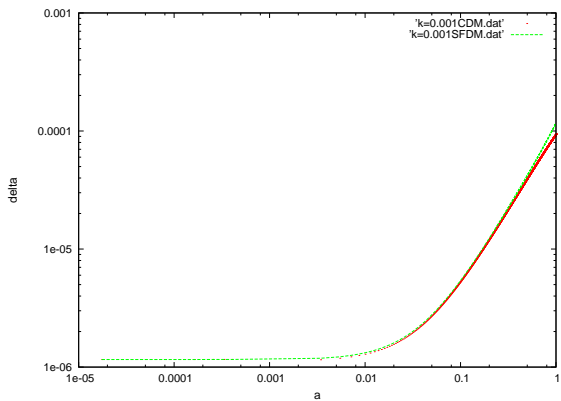}}
\caption{Evolution of the perturbations for the CDM model (dots) and SFDM model (lines) for $k=1*10^{-3}hMpc^{-1}$. Notice how after the 
epoch of equality ($a_{eq}\sim 10^{-4}$) the evolution of both perturbations in nearly identical, $a=1$ today. In this case we have taken $\lambda=0$.}
 \label{fig2}
 \end{figure}

 Now lets see what happens to the SFDM at this epoch ($a\geqslant a_{eq}$). The evolution of the perturbations in this case will 
be given by equation (\ref{delta}).

 In general we have that in equation (\ref{delta}) the term $v_q$ is very small throughout the evolution of the pertubations ($v_q\leq 
10^{-3}ms^{-1}$ for small $k$), so it really does not have a significant contribution on its evolution.
 
 When the condition $\lambda=0$ is taken we can have a BEC that might be or might not be stable, if there exists stability the results of 
SFDM are consistent with those obtained from CDM (in this case both equations (\ref{delta}) and (\ref{deltaCDM}) are almost equal), the condition of stability for the BEC in the SFDM case will come from the study of $\lambda$ together with $Q$. 
\begin{equation}
    \frac{\rmn{d}^2\delta}{\rmn{d}t^2}+2H\frac{\rmn{d}\delta}{\rmn{d}t}+\left(v_\rmn{q}^2\frac{k^2}{a^2}
    -4\pi G{\rho}_0\right)\delta=0,
   \label{deltal}
   \end{equation}

 As we can see in Fig. \ref{fig2} the perturbations used for the $\Lambda$CDM model grow in a similar way for the SFDM model, 
when $\lambda=0$, in this case both perturbations can give birth to structures quite similar in size, and this will happen with all the fluctuations as long as $k$ is kept small.

\begin{figure}
 \scalebox{0.6}{\includegraphics{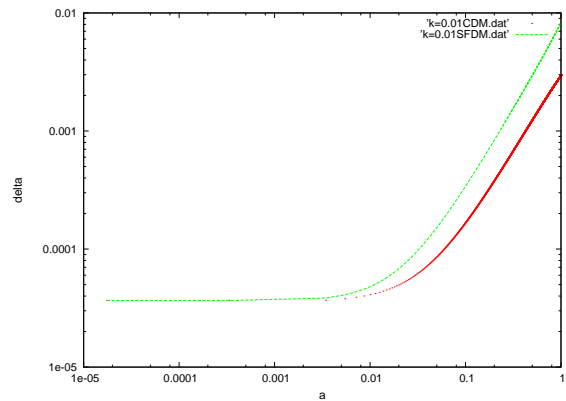}}
  \caption{Evolution of the perturbations for the CDM model (dots) and SFDM model (lines) for $k=1*10^{-2}hMpc^{-1}$ and $\lambda\neq 0$ 
and negative. Notice how after the epoch of equality ($a_{eq}\sim 10^{-4}$) the evolution of both perturbations is now different from the one in Fig. \ref{fig2}, $a=1$ today. In this case we can clearly see that the SFDM fluctuations grow quicker than those for the CDM model.}
 \label{fig3}
 \end{figure}

 When $\lambda\neq 0$ the results are quite different, so when discussing the evolution of the density perturbations, there are two 
different cases: i) In the case of $\lambda >0$ the amplitude of the density contrast tends to decrease as $\lambda$ grows bigger and bigger away from zero until the amplitude of the density takes negative values (around $\lambda\sim 10^{8}$), telling us that this kind of fluctuations can not grow in time, and hence do not form a BEC. ii) On the other hand if $\lambda <0$ the fluctuations for the density contrast alway grow despite their size, this results means either than the fluctuations grow and form a stable BEC or than the density grows because it is collapsing into a single 
point and our BEC might be unstable, the study of the stability of this fluctuations needs then to be studied with non-linear perturbation theory. These results are shown in Fig. \ref{fig3}, in both figures \ref{fig2} and \ref{fig3} the initial condition for $\delta$ goes as $\delta\sim 1*10^{-5}$ in accordance with the data obtained from WMAP.

If these fluctuations result stable and because they are big in size, this means that they can only give birth to large structures. These fluctuations can then help for the formation of large clusters or other large-scale structure in the Universe at its early stages (around $a\geqslant a_{eq}$). Then, as these kind of SFDM can only interact with radiation in a gravitational form it is not limited by its interaction with radiation, and the dark matter halos can then create potential wells that will collapse early in time giving enough time for the structures to form. Then if DM is some kind of SFDM, the luminous matter will follow the DM potentials giving birth to large-scale structure.

\section{Conclusions}

The new observational instruments and telescopes until today have perceived objects as far as $z=8.6$, \cite{b7}. The cosmic background 
radiation can bring us information from $z=1000$ to $z=2000$. But jet we can not see anything from the intermediate region, now we know of a possible galaxy that might be found at a distance of $z=10.56$ but it has jet to be confirmed.

 As seen earlier, as expected for the CDM model we obtained that for the matter dominated era the low-$k$ modes grow, when CDM decouples 
from radiation in a time just before recombination it grows in a milder way than it does in the matter dominated era (Fig. \ref{fig2}).

 Although in general a scalar field is not a fluid, it can be treated as if it behaved like one.The evolution of its density can be the 
appropriate for the purpose of structure formation, because locations with a high density of dark matter can support the formation of galactic structure.

 In this work we have assumed that there is only one component to the mass density, and that this component is given by the 
scalar field dark matter. In this case equation (\ref{delta}) is valid for all sub-horizon sized perturbations in our non-relativistic specie, so for sub-horizon perturbations the newtonian treatment worked with in the evolution of the perturbations suffices.

 The SFDM has provided to be an alternative model for the dark matter nature of the Universe. We have shown that the scalar 
field with an ultralight mass of $10^{-22}$ eV simulates the behavior of CDM in a Universe dominated by matter when $\lambda=0$, because in general in a matter dominated Universe for low-$k$, $v_\rmn{q}$ tends to be a very small quantity tending to zero, so from (\ref{delta}) we can see that on this era we will have the CDM profile given by (\ref{deltaCDM}), i.e., the SFDM density contrast profile is very similar to that of the $\Lambda$CDM model, Fig. \ref{fig2}. On the contrary for $\lambda\neq 0$ both models have different behavior as we can see from Fig. \ref{fig3}, results which show that linear fluctuations on the SFDM can grow in comparison with those of CDM, even at early times when the large-scale modes (small $k$) have entered the horizon just after $a_{eq}\sim 10^{-4}$, when it has decoupled from radiation, so the amplitudes of the density contrast start to grow faster than those for CDM around $a\sim 10^{-2}$. Here an important point is that although CDM can grow it does so in a hierarchical way, while from Fig. \ref{fig3} we can see that SFDM can have bigger fluctuations just before the $\Lambda$CDM model does, i.e., it might be that no hierarchical model of structure formation is needed for SFDM and is expected that for the non-linear fluctuations the behaviour will be quite the same as soon as the scalar field condensates, in a very early epoch when the energy of the Universe was about $\sim$ TeV. These facts can be the crucial difference between both models.

 As mentioned before, recent observations have taken us to very early epochs in the origin of the Universe, and have made us think that 
structure had already been formed, corresponding to $z\approx 7$.  It is clear from Fig. \ref{fig3} that at recombination $z\approx 1300$ there already existed defined perturbations in the energy density for the SFDM model, which can contribute to the early formation of structure. Then, if clusters could be formed as early as these z's, this would imply that $\Phi^2+\lambda\Phi^4$ as a model for dark matter could give an explanation for the characteristic masses that are being observed, and therefore it could solve some of the problems present in the standard $\Lambda$CDM model.

   Although the observational evidence seems to be in favor of some kind of cold dark matter, the last word has not been said. Astronomers 
hope to send satellites that will detect the finest details of the cosmic background radiation, which will help us to get information of structure at the time of recombination, from which it will be possible to deduce its evolution until now a days.

\section*{Acknowledgments}

This work was partially supported by CONACyT M\'exico, Instituto Avanzado de Cosmologia (IAC) collaboration. A. Su\'arez is supported by a CONACYT scholarships.

\bsp

\label{lastpage}


\begin{thebibliography}{99}

\bibitem[\protect\citeauthoryear{Alcubierre et al.}{2002}]{b1} Alcubierre M., Guzm\'an F. S., Matos T., N\'u\~nez D., Ure\~na L. A.,     
         Wiederhold P., 2002, Class Quant Grav, 19, 5017
\bibitem[\protect\citeauthoryear{Bardeen}{1980}]{b23} Bardeen J.M., 1980, Phys Rev D, 22, 1882
\bibitem[\protect\citeauthoryear{Bergman}{1992}]{b27} Bergman O., 1992, Phys Rev D, 46, 5474
\bibitem[\protect\citeauthoryear{Bernal, Matos \& N\'u\~nez}{2008}]{b2} Bernal A., Matos T., N\'u\~nez D., 2008, Rev. Mex. A.A., 44, 149
\bibitem[\protect\citeauthoryear{Boehmer \& Harko}{2007}]{b3} Boehmer C. G., Harko T., 2007, JCAP, 0706, 025
\bibitem[\protect\citeauthoryear{Bohm}{1952}]{b24} Bohm D., 1952, Phys Rev, 85, 180
\bibitem[\protect\citeauthoryear{Brilliantov \& P$\ddot{o}$schel}{2004}]{b31} Brilliantov N. V., P$\ddot{o}$schel T., 2004,  Kinetc Theory   
         of Granular Gases, Oxford University Press
\bibitem[\protect\citeauthoryear{Briscese}{2011}]{b22} Briscese F., 2011, Phys Lett B, 696, 315
\bibitem[\protect\citeauthoryear{Chaikin \& Lubensky}{1995}]{b30} Chaikin P.M., Lubensky T. C., 1995, Principles of condensed matter   
         physics, Cambridge University Press
\bibitem[\protect\citeauthoryear{Chiueh}{1998}]{b26} Chiueh T., 1998, Phys Rev E, 57, 4150
\bibitem[\protect\citeauthoryear{Clowe}{2006}]{b4} Clowe D. et al., 2006, ApJ, 648, 2
\bibitem[\protect\citeauthoryear{Dodelson}{2003}]{b32} Dodelson S., 2003, Modern Cosmology, Academic Press
\bibitem[\protect\citeauthoryear{Ginzburg \& Landau}{1985}]{b5} Ginzburg V. L., Landau L. D., 1950, Zh Eksp Teor Fiz, 20, 1064
\bibitem[\protect\citeauthoryear{Godr\'eche \& Manneville}{1998}]{b38} Godr\'eche C., Manneville P., 1998, Hydrodynamics and Nonlinear 
         Instabilities, Cambridge University Press
\bibitem[\protect\citeauthoryear{Gorini et al.}{2008}]{b50} Gorini V., Kamenshchik A. Y., Moschella U., Piattella O. F., Starobinsky A. A.,           2008, JCAP, 0802, 016
\bibitem[\protect\citeauthoryear{Hu, Barkana \& Gruzinov}{2008}]{b6} Hu W., Barkana R., Gruzinov A., 2008, preprint (astro-ph/0003365v2)
\bibitem[\protect\citeauthoryear{Huang}{1963}]{b34} Huang K., 1963, Statistical Mechanics, John Wiley \& Sons
\bibitem[\protect\citeauthoryear{Landau \& Lifshitz}{1980}]{b41} Landau L. D., Lifshitz E. M., 1980, Statistical Physics, Pergamon Press
\bibitem[\protect\citeauthoryear{Lee \& Koh}{1996}]{b20} Lee J., Koh I., 1996, Phys Rev D, 53, 2236
\bibitem[\protect\citeauthoryear{Lehnert et al.}{2010}]{b7} Lehnert M. D. et al., 2010, Nat, 467, 940
\bibitem[\protect\citeauthoryear{Liboff}{1980}]{b40} Liboff R. L., 1980, Introductory Quantum Mechanics, Addison-Wesley
\bibitem[\protect\citeauthoryear{Lundgren et al.}{2010}]{b25} Lundgren A. P., Bondarescu M., Bondarescu R., Balakrishna J., 2010, ApJ, 715,           L35
\bibitem[\protect\citeauthoryear{Ma \& Bertschinger}{1995}]{b19} Ma C.-P., E. Bertschinger, 1995, ApJ, 455, 7
\bibitem[\protect\citeauthoryear{Malik}{2009}]{b8} Malik K. A., 2009, Phys Rept, 475, 1
\bibitem[\protect\citeauthoryear{Matos}{2003}]{b28} Matos T., 2003, Rev. Mex. A.A., 49, 16
\bibitem[\protect\citeauthoryear{Matos \& Guzm\'an}{2000}]{b12} Matos T., Guzman F.S., 2000, Class Quant Grav, 17, L9
\bibitem[\protect\citeauthoryear{Matos, Maga\~na \& Su\'arez}{2010}]{b11} Matos T., Maga\~na J., Su\'arez A., 2010, The Open Astron Journal,          3, 94
\bibitem[\protect\citeauthoryear{Matos \& Ure\~na}{2000}]{b13} Matos T., Ure\~na L. A., 2000, Class Quant Grav, 17, L75
\bibitem[\protect\citeauthoryear{Matos \& Ure\~na}{2001}]{b10} Matos T., Ure\~na L. A., 2001, Phys Rev D, 63, 063506
\bibitem[\protect\citeauthoryear{Matos, V\'azquez-Gonz\'alez \& Maga\~na}{2009}]{b9} Matos T., V\'azquez-Gonz\'alez A.,Maga\~na J., 2009, 
         MNRAS, 389, 13957
\bibitem[\protect\citeauthoryear{Moore et al.}{1999}]{b16} Moore B., Ghigna S., Governato F., Lake G., Quinn T., Stadel J., Tozzi P., 1999,
         ApJ, 524, L19
\bibitem[\protect\citeauthoryear{$\ddot{O}$ttinger}{2005}]{b32} $\ddot{O}$ttinger H. C., 2005, Beyond Equilibrium Thermodynamics, John Wiley          \& Sons
\bibitem[\protect\citeauthoryear{Pathria}{1972}]{b39} Pathria R. K., 1972, Statistical Mechanics, Butterworth-Heinemann
\bibitem[\protect\citeauthoryear{Peebles \& Nusser}{2010}]{b14} Peebles P. J. E., Nusser A., 2010, Nat, 465, 565
\bibitem[\protect\citeauthoryear{Penny et al.}{2009}]{b15} Penny S. J., Conselice C. J., De Rijcke S., Held E. V.,  2009, MNRAS, 393, 1054
\bibitem[\protect\citeauthoryear{Pitaevskii \& Stringari}{2003}]{b29} Pitaevskii L., Stringari S., 2003, Bose-Einstein Condensation, Oxford           University Press
\bibitem[\protect\citeauthoryear{Reichl}{1998}]{b33} Reichl L.E., 1998, A Modern Course in Statistical Physics, John Wiley \& Sons
\bibitem[\protect\citeauthoryear{Sahni \& Wang}{2000}]{b21} Sahni V., Wang L., 2000, Phys Rev D, 62, 103527
\bibitem[\protect\citeauthoryear{Ure\~na}{2010}]{b17} Ure\~na L. A., 2010, AIP Conf Proc, 1318, 82
\bibitem[\protect\citeauthoryear{Woo \& Chiueh}{2008}]{b18} Woo T. P., Chiueh T., 2008, preprint (astro-ph/0806.0232v1)

\end{thebibliography}
\end{document}